\documentclass[12pt]{article}%
\usepackage{amsmath}
\usepackage{amsfonts}
\usepackage{amssymb}
\usepackage{graphicx}%
\begin{document}
\begin{titlepage}
\begin{flushright}
CAMS/05-01\\
\end{flushright}
\vspace{.3cm}
\begin{center}
\renewcommand{\thefootnote}{\fnsymbol{footnote}}
{\Large\bf Hermitian Geometry and Complex Space-Time} \vskip20mm
{\large\bf{Ali H.
Chamseddine \footnote{email: chams@aub.edu.lb}}}\\
\renewcommand{\thefootnote}{\arabic{footnote}}
\vskip2cm {\it Center for Advanced Mathematical Sciences (CAMS)
and\\
Physics Department, American University of Beirut, Lebanon.\\}
\end{center}
\vfill
\begin{center}
{\bf Abstract}
\end{center}
\vskip 1cm We consider a complex Hermitian manifold of complex
dimensions four with a Hermitian metric and a Chern connection. It
is shown that the action that determines the dynamics of the
metric is unique, provided that the linearized Einstein action
coupled to an antisymmetric tensor is obtained, in the limit when
the imaginary coordinates vanish. The unique action is of the
Chern-Simons type when expressed in terms of the K\"{a}hler form.
The antisymmetric tensor field has gauge transformations coming
from diffeomorphism invariance in the complex directions. The
equations of motion must be supplemented by boundary conditions
imposed on the Hermitian metric to give, in the limit of vanishing
imaginary coordinates, the low-energy effective action for a
curved metric coupled to an antisymmetric tensor.
\end{titlepage}

\section{\bigskip Introduction}

The idea of complexifying space-time in general relativity was put forward in
the early sixties. It appeared in different but related lines of research.
These include complexifying the four-dimensional manifold and equipping it
with a holomorphic metric, asymptotically complex null surfaces and theory of
twistors \cite{Synge}, \cite{Newman}, \cite{Penrose}, \cite{PR},
\cite{Plebanski}, \cite{Flaherty 76}, \cite{Flaherty 80}. More recently,
Witten \cite{Witten} considered string propagation on complexified space-time
where he presented some evidence that the imaginary part of the complex
coordinates enters in the study of the high-energy behavior of scattering
amplitudes \cite{GM}. In this string picture it is assumed that the imaginary
parts of the coordinates are small at low-energies. At a fundamental level the
complex coordinates $X^{\mu}$, $\mu=1,\cdots,d$ \ with complex conjugates
$\overline{X^{\mu}}\equiv X^{\overline{\mu}}$ are described by the topological
$\sigma$ model action \cite{Witten}
\[
I=%
%TCIMACRO{\dint }%
%BeginExpansion
{\displaystyle\int}
%EndExpansion
d\sigma d\overline{\sigma}g_{\mu\overline{\nu}}\left(  X(\sigma,\overline
{\sigma}),\overline{X}\left(  \sigma,\overline{\sigma}\right)  \right)
\partial_{\overline{\sigma}}X^{\mu}\partial_{\sigma}X^{\overline{\nu}},
\]
where the world-sheet coordinates are denoted by $\sigma$ and $\overline
{\sigma}$, and where the background metric for the complex d-dimensional
manifold M is Hermitian so that
\[
\overline{g_{\mu\overline{\nu}}}=g_{\nu\overline{\mu}},\qquad g_{\mu\nu
}=g_{\overline{\mu}\,\overline{\nu}}=0.
\]
Decomposing the metric into real and imaginary components%
\[
g_{\mu\overline{\nu}}=G_{\mu\nu}+iB_{\mu\nu},
\]
the hermiticity condition implies that $G_{\mu\nu}$ is symmetric and
$B_{\mu\nu}$ is antisymmetric. The low-energy effective string action is given
by the Einstein-Hilbert action coupled to the field strength of the
antisymmetric tensor. This can be related to the invariance of the sigma model
under complex transformations $X^{\mu}\rightarrow X^{\mu}+\zeta^{\mu}\left(
X\right)  ,$ $X^{\overline{\mu}}\rightarrow X^{\overline{\mu}}+\zeta
^{\overline{\mu}}\left(  \overline{X}\right)  $.

A related phenomena was observed in noncommutative geometry \cite{AC} where
the space-time coordinates are deformed and become noncommuting, $\left[
x^{\mu},x^{\nu}\right]  =i\theta^{\mu\nu}$ \cite{CDS}. Furthermore, It was
found that in the effective action of open-string theory, the inverse of the
combinations $\left(  G_{\mu\nu}+B_{\mu\nu}\right)  ^{-1}$ does appear
\cite{SW}. This was taken as a motivation to study the dynamics of a complex
Hermitian metric on a real manifold \cite{AHC}, \ considered first by Einstein
and Strauss \cite{Einstein}. In \cite{AHC} it was shown that the invariant
action constructed have the required behavior for the propagation of the
fields $G_{\mu\nu}$ and $B_{\mu\nu}$ at the linearized level, but problems do
arise when non-linear interactions are taken into account. This is due to the
fact that there is no gauge symmetry to prevent the ghost components of
$B_{\mu\nu}$ from propagating. It is then important to address the question of
whether it is possible to have consistent interactions in which the field
$B_{\mu\nu}$ appears explicitly in analogy with $G_{\mu\nu}$ and not only
through the combination of derivatives%
\[
H_{\mu\nu\rho}=\partial_{\mu}B_{\nu\rho}+\partial_{\nu}B_{\rho\mu}%
+\partial_{\rho}B_{\mu\nu}.
\]
This suggests that the gauge parameters for the transformation $B_{\mu\nu
}\rightarrow B_{\mu\nu}+\partial_{\mu}\Lambda_{\nu}-\partial_{\nu}\Lambda
_{\mu}$ that keep $H_{\mu\nu\rho}$ invariant must be combined with the
diffeomorphism parameters on the real manifold. For this to happen there must
be diffeomorphism invariance of the Hermitian manifold $M$ of complex
dimensions $d,$ with complex coordinates $z^{\mu}=x^{\mu}+iy^{\mu}$,
$\mu=1,\cdots,d.$ The line element is then given by \cite{Goldberg}%
\[
ds^{2}=2g_{\mu\overline{\nu}}dz^{\mu}d\overline{z}^{\overline{\nu}},
\]
where we have denoted $\overline{z^{\mu}}=z^{\overline{\mu}}.$ The metric
preserves its form under infinitesimal transformations%
\begin{align*}
z^{\mu} &  \rightarrow z^{\mu}-\zeta^{\mu}\left(  z\right)  ,\\
z^{\overline{\mu}} &  \rightarrow z^{\overline{\mu}}-\zeta^{\overline{\mu}%
}\left(  \overline{z}\right)  ,
\end{align*}
as can be seen from the transformations%
\begin{align*}
0 &  =\delta g_{\mu\nu}=\partial_{\mu}\zeta^{\overline{\lambda}}%
g_{\overline{\lambda}\nu}+\partial_{\nu}\zeta^{\overline{\lambda}}%
g_{\mu\overline{\lambda}},\\
0 &  =\delta g_{\overline{\mu}\,\overline{\nu}}=\partial_{\overline{\mu}}%
\zeta^{\lambda}g_{\lambda\overline{\nu}}+\partial_{\overline{\nu}}%
\zeta^{\lambda}g_{\overline{\mu}\lambda},\\
\delta g_{\mu\overline{\nu}} &  =\partial_{\mu}\zeta^{\lambda}g_{\lambda
\overline{\nu}}+\partial_{\overline{\nu}}\zeta^{\overline{\lambda}}%
g_{\mu\overline{\lambda}}+\zeta^{\lambda}\partial_{\lambda}g_{\mu\overline
{\nu}}+\zeta^{\overline{\lambda}}\partial_{\overline{\lambda}}g_{\mu
\overline{\nu}}.
\end{align*}
It is instructive to express these transformations in terms of the fields
$G_{\mu\nu}(x,y)$ and $B_{\mu\nu}(x,y)$ by writing
\begin{align*}
\zeta^{\mu}(z) &  =\alpha^{\mu}(x,y)+i\beta^{\mu}(x,y),\\
\zeta^{\overline{\mu}}(\overline{z}) &  =\alpha^{\mu}(x,y)-i\beta^{\mu}(x,y).
\end{align*}
The holomorphicity conditions on $\zeta^{\mu}$ and $\zeta^{\overline{\mu}}$
imply the relations
\begin{align*}
\partial_{\mu}^{y}\beta^{\nu} &  =\partial_{\mu}^{x}\alpha^{\nu},\\
\partial_{\mu}^{y}\alpha^{\nu} &  =-\partial_{\mu}^{x}\beta^{\nu},
\end{align*}
where we have denoted
\[
\partial_{\mu}^{y}=\frac{\partial}{\partial y^{\mu}},\qquad\partial_{\mu}%
^{x}=\frac{\partial}{\partial x^{\mu}}.
\]
The transformations of $G_{\mu\nu}(x,y)$ and $B_{\mu\nu}(x,y)$ are then given
by%
\begin{align*}
\delta G_{\mu\nu}(x,y) &  =\partial_{\mu}^{x}\alpha^{\lambda}G_{\lambda\nu
}+\partial_{\nu}^{x}\alpha^{\lambda}G_{\mu\lambda}+\alpha^{\lambda}%
\partial_{\lambda}^{x}G_{\mu\nu}\\
&  -\partial_{\mu}^{x}\beta^{\lambda}B_{\lambda\nu}+\partial_{\nu}^{x}%
\beta^{\lambda}B_{\mu\lambda}+\beta^{\lambda}\partial_{\lambda}^{y}G_{\mu\nu
},\\
\delta B_{\mu\nu}(x,y) &  =\partial_{\mu}^{x}\beta^{\lambda}G_{\lambda\nu
}-\partial_{\nu}^{x}\beta^{\lambda}G_{\mu\lambda}+\alpha^{\lambda}%
\partial_{\lambda}^{x}B_{\mu\nu}\\
&  +\partial_{\mu}^{x}\alpha^{\lambda}B_{\lambda\nu}+\partial_{\nu}^{x}%
\alpha^{\lambda}B_{\mu\lambda}+\beta^{\lambda}\partial_{\lambda}^{y}B_{\mu\nu
}.
\end{align*}
One readily recognizes that in the vicinity of small $y^{\mu}$ the fields
$G_{\mu\nu}(x,0)$ and $B_{\mu\nu}(x,0)$ transform as symmetric and
antisymmetric tensors with gauge parameters $\alpha^{\mu}(x)$ and $\beta^{\mu
}(x)$ where
\begin{align*}
\alpha^{\mu}(x,y) &  =\alpha^{\mu}(x)-\partial_{\nu}^{x}\beta^{\mu}(x)y^{\nu
}+O(y^{2}),\\
\beta^{\mu}(x,y) &  =\beta^{\mu}(x)+\partial_{\nu}^{x}\alpha^{\mu}(x)y^{\nu
}+O(y^{2}),
\end{align*}
as implied by the holomorphicity conditions.

The purpose of this work is to investigate the dynamics of the Hermitian
metric $g_{\mu\overline{\nu}}$ on a complex space-time with complex dimensions
four, such that in the limit of vanishing imaginary values of the coordinates,
the action reduces to that of a symmetric metric $G_{\mu\nu}$ and an
antisymmetric field $B_{\mu\nu}.$ The plan of this paper is as follows. In
section two we summarize the essentials of Hermitian geometry. In section
three we construct the most general action which gives, in the linearized
limit, the correct equations of motion for a symmetric metric $G_{\mu\nu}$ and
an antisymmetric field $B_{\mu\nu}$ and show that the action is unique. In
section four we impose constraints on the torsion and curvature in the four
dimensional limit where the imaginary values of the coordinates vanish and
study the equations of motion . Section five is the conclusion.

\section{\bigskip Hermitian Geometry}

The Hermitian manifold $M$ of complex dimensions $d$ \ is defined as a
Riemannian manifold with real dimensions $2d$ with Riemannian metric $g_{ij}$
and complex coordinates $z^{i}=\left\{  z^{\mu},z^{\overline{\mu}}\right\}  $
where Latin indices $i,j,k,\cdots,$ run over the range $1,2,\cdots
,d,\overline{1},\overline{2},\cdots,\overline{d}.$ The invariant line element
is then \cite{Yano}
\[
ds^{2}=g_{ij}dz^{i}dz^{j},
\]
where the metric $g_{ij}$ is hybrid%
\[
g_{ij}=\left(
\begin{array}
[c]{cc}%
0 & g_{\mu\overline{\nu}}\\
g_{\nu\overline{\mu}} & 0
\end{array}
\right)  .
\]
It has also an integrable complex structure $F_{i}^{j}$ satisfying
\[
F_{i}^{k}F_{k}^{j}=-\delta_{i}^{j},
\]
and with a vanishing Nijenhuis tensor
\[
N_{ji}^{\hspace{0.06in}h}=F_{j}^{t}\left(  \partial_{t}F_{i}^{h}-\partial
_{i}F_{t}^{h}\right)  -F_{i}^{t}\left(  \partial_{t}F_{j}^{h}-\partial
_{j}F_{t}^{h}\right)  .
\]
The complex structure has components
\[
F_{i}^{j}=\left(
\begin{array}
[c]{cc}%
i\delta_{\mu}^{\nu} & 0\\
0 & -i\delta_{\overline{\mu}}^{\overline{\nu}}%
\end{array}
\right)  .
\]
The affine connection with torsion $\Gamma_{ij}^{h}$ is introduced so that the
following two conditions are satisfied%
\begin{align*}
\nabla_{k}g_{ij}  &  =\partial_{k}g_{ij}-\Gamma_{ik}^{h}g_{hj}-\Gamma_{jk}%
^{h}g_{ih}=0,\\
\nabla_{k}F_{i}^{j}  &  =\partial_{k}F_{i}^{j}-\Gamma_{ik}^{h}F_{h}^{j}%
+\Gamma_{hk}^{j}F_{i}^{h}=0.
\end{align*}
These conditions do not determine the affine connection uniquely and there
exists several possibilities used in the literature. We shall adopt the Chern
connection, which is the one most commonly used, . It is defined by
prescribing that the $(2d)^{2}$ linear differential forms
\[
\omega_{\;j}^{i}=\Gamma_{jk}^{i}dz^{k},
\]
be such that $\omega_{\;\nu}^{\mu}$ and $\omega_{\;\overline{\nu}}%
^{\overline{\mu}}$ are given by \cite{Goldberg}
\begin{align*}
\omega_{\;\nu}^{\mu}  &  =\Gamma_{\nu\rho}^{\mu}dz^{\rho},\\
\overline{\omega_{\;\nu}^{\mu}}  &  =\omega_{\;\overline{\nu}}^{\overline{\mu
}}=\Gamma_{\overline{\nu}\,\overline{\rho}}^{\overline{\mu}}dz^{\overline
{\rho}},
\end{align*}
with the remaining $(2d)^{2}$ forms set equal to zero. For $\omega_{\;\nu
}^{\mu}$ to have a metrical connection the differential of the metric tensor
$g$ must be given by
\[
dg_{\mu\overline{\nu}}=\omega_{\;\mu}^{\rho}g_{\rho\overline{\nu}}%
+\omega_{\;\overline{\nu}}^{\overline{\rho}}g_{\mu\overline{\rho}},
\]
from which we obtain%
\[
\partial_{\lambda}g_{\mu\overline{\nu}}dz^{\lambda}+\partial_{\overline
{\lambda}}g_{\mu\overline{\nu}}dz^{\overline{\lambda}}=\Gamma_{\mu\lambda
}^{\rho}g_{\rho\overline{\nu}}dz^{\lambda}+\Gamma_{\overline{\nu}%
\overline{\lambda}}^{\overline{\rho}}g_{\mu\overline{\rho}}dz^{\overline
{\lambda}},
\]
so that
\begin{align*}
\Gamma_{\mu\lambda}^{\rho}  &  =g^{\overline{\nu}\rho}\partial_{\lambda}%
g_{\mu\overline{\nu}},\\
\Gamma_{\overline{\nu}\overline{\lambda}}^{\overline{\rho}}  &  =g^{\overline
{\rho}\mu}\partial_{\overline{\lambda}}g_{\mu\overline{\nu}},
\end{align*}
where the inverse metric $g^{\overline{\nu}\mu}$ is defined by
\[
g^{\overline{\nu}\mu}g_{\mu\overline{\kappa}}=\delta_{\overline{\kappa}%
}^{\overline{\nu}}.
\]
The condition $\nabla_{k}F_{i}^{j}=0$ is then automatically satisfied and the
connection is metric. The torsion forms are defined by
\begin{align*}
\Theta^{\mu}  &  \equiv-\frac{1}{2}T_{\nu\rho}^{\hspace{0.1in}\mu}dz^{\nu
}\wedge dz^{\rho}\\
&  =\omega_{\;\nu}^{\mu}dz^{\nu}=-\Gamma_{\nu\rho}^{\mu}dz^{\nu}\wedge
dz^{\rho},
\end{align*}
which implies that%
\begin{align*}
T_{\nu\rho}^{\hspace{0.1in}\mu}  &  =\Gamma_{\nu\rho}^{\mu}-\Gamma_{\rho\nu
}^{\mu}\\
&  =g^{\overline{\sigma}\mu}\left(  \partial_{\rho}g_{\nu\overline{\sigma}%
}-\partial_{\nu}g_{\rho\overline{\sigma}}\right)  .
\end{align*}
The torsion form is related to the differential of the Hermitian form
\[
F=\frac{1}{2}F_{ij}dz^{i}\wedge dz^{j},
\]
where
\[
F_{ij}=F_{i}^{k}g_{kj}=-F_{ji},
\]
is antisymmetric and satisfy
\begin{align*}
F_{\mu\nu}  &  =0=F_{\overline{\mu}\,\overline{\nu}},\\
F_{\mu\overline{\nu}}  &  =ig_{\mu\overline{\nu}}=-F_{\overline{\nu}\mu},
\end{align*}
so that
\[
F=ig_{\mu\overline{\nu}}dz^{\mu}\wedge dz^{\overline{\nu}}.
\]
The differential of $F$ is then%
\[
dF=\frac{1}{6}F_{ijk}dz^{i}\wedge dz^{j}\wedge dz^{k},
\]
so that
\[
F_{ijk}=\partial_{i}F_{jk}+\partial_{j}F_{ki}+\partial_{k}F_{ij}.
\]
The only non-vanishing components of this tensor are
\begin{align*}
F_{\mu\nu\overline{\rho}}  &  =i\left(  \partial_{\mu}g_{\nu\overline{\rho}%
}-\partial_{\nu}g_{\mu\overline{\rho}}\right)  =-iT_{\mu\nu}^{\hspace
{0.1in}\sigma}g_{\sigma\overline{\rho}}=-iT_{\mu\nu\overline{\rho}},\\
F_{\overline{\mu}\,\overline{\nu}\rho}  &  =-i\left(  \partial_{\overline{\mu
}}g_{\rho\overline{\nu}}-\partial_{\overline{\nu}}g_{\rho\overline{\mu}%
}\right)  =iT_{\overline{\mu}\,\overline{\nu}}^{\hspace{0.1in}\overline
{\sigma}}g_{\rho\overline{\sigma}}=iT_{\overline{\mu}\,\overline{\nu}\rho}.
\end{align*}
The curvature tensor of the metric connection is constructed in the usual
manner
\[
\Omega_{\;j}^{i}=d\omega_{\;j}^{i}-\omega_{\;k}^{i}\wedge\omega_{\;j}^{k},
\]
with the only non-vanishing components $\Omega_{\;\mu}^{\nu}$ and
$\Omega_{\;\overline{\mu}}^{\overline{\nu}}.$ These are given by
\begin{align*}
\Omega_{\;\mu}^{\nu}  &  =-R_{\;\mu\kappa\lambda}^{\nu}dz^{\kappa}\wedge
dz^{\lambda}-R_{\;\mu\kappa\overline{\lambda}}^{\nu}dz^{\kappa}\wedge
dz^{\overline{\lambda}}\\
&  =\left(  \partial_{\kappa}\Gamma_{\mu\lambda}^{\nu}-\Gamma_{\mu\kappa
}^{\rho}\Gamma_{\rho\lambda}^{\nu}\right)  dz^{\kappa}\wedge dz^{\lambda
}-\partial_{\overline{\lambda}}\Gamma_{\mu\kappa}^{\nu}dz^{\kappa}\wedge
dz^{\overline{\lambda}}.
\end{align*}
Comparing both sides we obtain%
\begin{align*}
R_{\;\mu\kappa\lambda}^{\nu}  &  =\partial_{\lambda}\Gamma_{\mu\kappa}^{\nu
}-\partial_{\kappa}\Gamma_{\mu\lambda}^{\nu}+\Gamma_{\mu\kappa}^{\rho}%
\Gamma_{\rho\lambda}^{\nu}-\Gamma_{\mu\lambda}^{\rho}\Gamma_{\rho\kappa}^{\nu
},\\
R_{\;\mu\kappa\overline{\lambda}}^{\nu}  &  =\partial_{\overline{\lambda}%
}\Gamma_{\mu\kappa}^{\nu}.
\end{align*}
One can easily show that%
\begin{align*}
R_{\;\mu\kappa\lambda}^{\nu}  &  =0,\\
R_{\;\mu\kappa\overline{\lambda}}^{\nu}  &  =g^{\overline{\rho}\nu}%
\partial_{\kappa}\partial_{\overline{\lambda}}g_{\mu\overline{\rho}}%
+\partial_{\overline{\lambda}}g^{\overline{\rho}\nu}\partial_{\kappa}%
g_{\mu\overline{\rho}}.
\end{align*}
Transvecting the last relation with $g_{\nu\overline{\sigma}}$ we obtain
\[
-R_{\mu\overline{\sigma}\kappa\overline{\lambda}}=\partial_{\kappa}%
\partial_{\overline{\lambda}}g_{\mu\overline{\sigma}}+g_{\nu\overline{\sigma}%
}\partial_{\overline{\lambda}}g^{\overline{\rho}\nu}\partial_{\kappa}%
g_{\mu\overline{\rho}}.
\]
Therefore the only non-vanishing covariant components of the curvature tensor
are
\[
R_{\mu\overline{\nu}\kappa\overline{\lambda}},\quad R_{\mu\overline{\nu
\,}\,\overline{\kappa}\lambda},\quad R_{\overline{\mu}\nu\kappa\overline
{\lambda}},\quad R_{\overline{\mu}\nu\overline{\kappa}\lambda},
\]
which are related by%
\[
R_{\mu\overline{\nu}\kappa\overline{\lambda}}=-R_{\overline{\nu}\mu
\kappa\overline{\lambda}}=-R_{\mu\overline{\nu}\overline{\lambda}\kappa},
\]
and satisfy the first Bianchi identity \cite{Goldberg}%
\[
R_{\;\mu\kappa\overline{\lambda}}^{\nu}-R_{\;\kappa\mu\overline{\lambda}}%
^{\nu}=\nabla_{\overline{\lambda}}T_{\mu\kappa}^{\quad\nu}.
\]
The second Bianchi identity is given by%
\[
\nabla_{\rho}R_{\mu\overline{\nu}\kappa\overline{\lambda}}-\nabla_{\kappa
}R_{\mu\overline{\nu}\rho\overline{\lambda}}=R_{\mu\overline{\nu}%
\sigma\overline{\lambda}}T_{\rho\kappa}^{\quad\sigma},
\]
together with the conjugate relations. There are three possible contractions
for the curvature tensor which are called the Ricci tensors
\[
R_{\mu\overline{\nu}}=-g^{\overline{\lambda}\kappa}R_{\mu\overline{\lambda
}\kappa\overline{\nu}},\quad S_{\mu\overline{\nu}}=-g^{\overline{\lambda
}\kappa}R_{\mu\overline{\nu}\kappa\overline{\lambda}},\quad T_{\mu
\overline{\nu}}=-g^{\overline{\lambda}\kappa}R_{\kappa\overline{\lambda}%
\mu\overline{\nu}}.
\]
Upon further contraction these result in two possible curvature scalars%
\[
R=g^{\overline{\nu}\mu}R_{\mu\overline{\nu}},\quad S=g^{\overline{\nu}\mu
}S_{\mu\overline{\nu}}=g^{\overline{\nu}\mu}T_{\mu\overline{\nu}}.
\]
Note that when the torsion tensors vanishes, the manifold $M$ \ becomes
K\"{a}hler. We shall not impose the K\"{a}hler condition as we are interested
in Hermitian non-K\"{a}hlerian geometry. We note that it is also possible to
consider the Levi-Civita connection $\mathring{\Gamma}_{ij}^{k}$ and the
associated Riemann curvature $K_{kij}^{\hspace{0.13in}h}$ where
\begin{align*}
\mathring{\Gamma}_{ij}^{k}  &  =\frac{1}{2}g^{kl}\left(  \partial_{i}%
g_{lj}+\partial_{j}g_{il}-\partial_{l}g_{ij}\right)  ,\\
K_{kij}^{\hspace{0.13in}\,h}  &  =\partial_{k}\mathring{\Gamma}_{ij}%
^{h}-\partial_{i}\mathring{\Gamma}_{kj}^{h}+\mathring{\Gamma}_{kt}%
^{h}\mathring{\Gamma}_{ij}^{t}-\mathring{\Gamma}_{it}^{h}\mathring{\Gamma
}_{kj}^{t}.
\end{align*}
The relation between the Chern connection and the Levi-Civita connection is
given by%
\[
\Gamma_{ij}^{k}=\mathring{\Gamma}_{ij}^{k}+\frac{1}{2}\left(  T_{ij}%
^{\hspace{0.07in}k}-T_{\hspace{0.03in}\,ij}^{k}-T_{\hspace{0.03in}\,ji}%
^{k}\right)  .
\]
The Ricci tensor and curvature scalar are $K_{ij}=$ $K_{tij}^{\hspace
{0.13in}t}$ and $K=g^{ij}K_{ij}.$ Moreover $H_{kj}=K_{kji}^{\hspace
{0.13in}\,t}F_{t}^{i}$ and $H=g^{kj}H_{kj}.$ The two scalar curvatures $K$ and
$H$ are related by \cite{Gaud}
\[
K-H=\mathring{\nabla}^{h}F^{ij}\mathring{\nabla}_{j}F_{ih}-\mathring{\nabla
}^{k}F_{ki}\mathring{\nabla}_{h}F^{hi}-2F^{ji}\mathring{\nabla}_{j}%
\mathring{\nabla}^{k}F_{ki}.
\]
There are also relations between curvatures of the Chern connection and those
of the Levi-Civita connection, mainly \cite{Gaud}%
\[
\frac{1}{2}K=S-\nabla^{\mu}T_{\mu}-\nabla^{\overline{\mu}}T_{\overline{\mu}%
}-T_{\mu}T_{\overline{\nu}}g^{\overline{\nu}\mu},
\]
where $T_{\mu}=T_{\mu\nu}^{\hspace{0.1in}\nu}.$ There are two natural
conditions that can be imposed on the torsion. The first is $T_{\mu}=0$ which
results in a semi-K\"{a}hler manifold. The other is when the torsion is
complex analytic so that $\nabla_{\overline{\lambda}}T_{\mu\kappa}%
^{\hspace{0.1in}\nu}=0$ implying that the curvature tensor has the same
symmetry properties as in the K\"{a}hler case. In this work we shall not
impose any conditions on the torsion tensor.

\section{\bigskip An invariant action}

We now specialize to the realistic case of a complexified four dimensional
space-time. To construct invariants up to second order in derivatives we write
the following possible terms%
\[
I=%
%TCIMACRO{\dint \limits_{M^{4}}}%
%BeginExpansion
{\displaystyle\int\limits_{M^{4}}}
%EndExpansion
d^{4}zd^{4}\overline{z}g\left(  aR+bS+c\,T_{\mu\nu\overline{\kappa}%
}T_{\overline{\rho}\,\overline{\sigma}\lambda}g^{\overline{\rho}\mu
}g^{\overline{\sigma}\nu}g^{\overline{\kappa}\lambda}+d\,T_{\mu\nu
\overline{\kappa}}T_{\overline{\rho}\,\overline{\sigma}\lambda}g^{\overline
{\rho}\mu}g^{\overline{\sigma}\lambda}g^{\overline{\kappa}\nu}+e\right)  .
\]
The density factor is $\left\vert \det g_{ij}\right\vert ^{\frac{1}{2}}=\det
g_{\mu\overline{\nu}}\equiv g.$ We shall set the cosmological term to zero
$\left(  e=0\right)  .$ The above action can equivalently be written in terms
of the Riemannian metric $g_{ij}$ in the form
\[
I=%
%TCIMACRO{\dint \limits_{M}}%
%BeginExpansion
{\displaystyle\int\limits_{M}}
%EndExpansion
d^{4}zd^{4}\overline{z}\left\vert \det g_{ij}\right\vert ^{\frac{1}{2}}\left(
a^{\prime}K+b^{\prime}H+c^{\prime}F_{ijk}F^{ijk}+d^{\prime}F_{i}F^{i}\right)
,
\]
where $F_{i}=F_{ijk}F^{jk}$ and $a^{\prime},b^{\prime},c^{\prime},d^{\prime}$
are parameters linearly related to the parameters $a,b,c,d.$ We shall now
impose the requirement that the linearized action, in the limit of
$y\rightarrow0$ gives the correct kinetic terms for $G_{\mu\nu}(x)$ and
$B_{\mu\nu}(x).$ Therefore writing
\begin{align*}
G_{\mu\nu}\left(  x,y\right)   &  =\eta_{\mu\nu}+h_{\mu\nu}(x),\\
B_{\mu\nu}(x,y) &  =B_{\mu\nu}(x),
\end{align*}
and keeping only quadratic terms in the action, we obtain, after integrating
by parts, the quadratic $h_{\mu\nu}$ terms,%
\begin{align*}
I &  =%
%TCIMACRO{\dint }%
%BeginExpansion
{\displaystyle\int}
%EndExpansion
d^{4}xd^{4}y\left(  2c\partial_{\kappa}^{x}h_{\mu\nu}\partial^{x\kappa}%
h_{\mu\nu}+\left(  a-2c+d\right)  \partial^{x\nu}h_{\mu\nu}\partial_{\lambda
}^{x}h^{\mu\lambda}+\right.  \\
&  \hspace{0.5in}\left.  -\left(  a-b+2d\right)  \partial^{x\nu}h_{\mu\nu
}\partial^{x\nu}h_{\lambda}^{\hspace{0.05in}\lambda}+\left(  d-b\right)
\partial_{\mu}^{x}h_{\nu}^{\hspace{0.05in}\nu}\partial^{x\mu}h_{\lambda
}^{\hspace{0.05in}\lambda}\right)  .
\end{align*}
Comparing with the linearized Einstein action we obtain the following
conditions
\[
2c=1,\quad a-2c+d=-2,\quad-a+b-2d=2,\quad d-b=-1,
\]
which are equivalent to
\[
b=-a,\quad c=\frac{1}{2},\quad d=-1-a.
\]
With this choice of coefficients, the quadratic $B$ contributions simplify to
\[%
%TCIMACRO{\dint }%
%BeginExpansion
{\displaystyle\int}
%EndExpansion
d^{4}xd^{4}y\left(  \partial_{\mu}^{x}B_{\nu\rho}\partial^{x\mu}B^{\nu\rho
}-2\partial^{x\mu}B_{\mu\lambda}\partial_{\nu}^{x}B^{\nu\lambda}\right)  ,
\]
which is identical to the term
\[
\frac{1}{3}%
%TCIMACRO{\dint }%
%BeginExpansion
{\displaystyle\int}
%EndExpansion
d^{4}xd^{4}yH_{\mu\nu\rho}H^{\mu\nu\rho},
\]
where $H_{\mu\nu\rho}=\partial_{\mu}^{x}B_{\nu\rho}+\partial_{\nu}^{x}%
B_{\rho\mu}+\partial_{\rho}^{x}B_{\mu\nu}.$ The action can then be regrouped
into the form%
\begin{align*}
I &  =%
%TCIMACRO{\dint \limits_{M}}%
%BeginExpansion
{\displaystyle\int\limits_{M}}
%EndExpansion
d^{4}zd^{4}\overline{z}g\left(  a\left(  R-S-T_{\mu\nu\overline{\kappa}%
}T_{\overline{\rho}\,\overline{\sigma}\lambda}g^{\overline{\rho}\mu
}g^{\overline{\sigma}\lambda}g^{\overline{\kappa}\nu}\right)  \right.  \\
&  +\left.  \frac{1}{2}T_{\mu\nu\overline{\kappa}}T_{\overline{\rho
}\,\overline{\sigma}\lambda}\left(  g^{\overline{\rho}\mu}g^{\overline{\sigma
}\nu}g^{\overline{\kappa}\lambda}-2g^{\overline{\rho}\mu}g^{\overline{\sigma
}\lambda}g^{\overline{\kappa}\nu}\right)  \right)  .
\end{align*}
Using the first Bianchi identity we have
\begin{align*}%
%TCIMACRO{\dint \limits_{M}}%
%BeginExpansion
{\displaystyle\int\limits_{M}}
%EndExpansion
d^{4}zd^{4}\overline{z}g\left(  R-S\right)   &  =%
%TCIMACRO{\dint \limits_{M}}%
%BeginExpansion
{\displaystyle\int\limits_{M}}
%EndExpansion
d^{4}zd^{4}\overline{z}gg^{\overline{\lambda}\mu}\partial_{\overline{\lambda}%
}T_{\mu\nu}^{\quad\nu}\\
&  =%
%TCIMACRO{\dint \limits_{M}}%
%BeginExpansion
{\displaystyle\int\limits_{M}}
%EndExpansion
d^{4}zd^{4}\overline{z}gT_{\mu\nu\overline{\kappa}}T_{\overline{\rho
}\,\overline{\sigma}\lambda}g^{\overline{\rho}\mu}g^{\overline{\sigma}\lambda
}g^{\overline{\kappa}\nu},
\end{align*}
where we have integrated by parts and ignored a surface term. This imply that
the group of terms with coefficient $a$ drop out, and the action becomes
unique:%
\[
I=\frac{1}{2}%
%TCIMACRO{\dint \limits_{M}}%
%BeginExpansion
{\displaystyle\int\limits_{M}}
%EndExpansion
d^{4}zd^{4}\overline{z}gT_{\mu\nu\overline{\kappa}}T_{\overline{\rho
\,}\overline{\sigma}\lambda}\left(  g^{\overline{\rho}\mu}g^{\overline{\sigma
}\nu}g^{\overline{\kappa}\lambda}-2g^{\overline{\rho}\mu}g^{\overline{\sigma
}\lambda}g^{\overline{\kappa}\nu}\right)  .
\]
Substituting for the torsion tensor in terms of the metric $g_{\mu
\overline{\nu}}$, the above action reduces to
\[
I=\frac{1}{2}%
%TCIMACRO{\dint \limits_{M}}%
%BeginExpansion
{\displaystyle\int\limits_{M}}
%EndExpansion
d^{4}zd^{4}\overline{z}gX^{\overline{\kappa}\overline{\lambda}\overline
{\sigma}\mu\nu\rho}\partial_{\nu}g_{\mu\overline{\sigma}}\partial
_{\overline{\lambda}}g_{\rho\overline{\kappa}},
\]
where
\begin{align*}
X^{\overline{\kappa}\overline{\lambda}\overline{\sigma}\mu\nu\rho} &
=g^{\overline{\sigma}\rho}\left(  g^{\overline{\kappa}\mu}g^{\overline
{\lambda}\nu}-g^{\overline{\kappa}\nu}g^{\overline{\lambda}\mu}\right)
+g^{\overline{\sigma}\mu}\left(  g^{\overline{\kappa}\nu}g^{\overline{\lambda
}\rho}-g^{\overline{\kappa}\rho}g^{\overline{\lambda}\nu}\right)  \\
&  +g^{\overline{\sigma}\nu}\left(  g^{\overline{\kappa}\rho}g^{\overline
{\lambda}\mu}-g^{\overline{\kappa}\mu}g^{\overline{\lambda}\rho}\right)  ,
\end{align*}
which is completely antisymmetric in the indices $\mu\nu\rho$ and in
$\overline{\kappa}\overline{\lambda}\overline{\sigma}$%
\[
X^{\overline{\kappa}\overline{\lambda}\overline{\sigma}\mu\nu\rho}=X^{\left[
\overline{\kappa}\overline{\lambda}\overline{\sigma}\right]  \left[  \mu
\nu\rho\right]  }.
\]
This is remarkable because the simple requirement that the linearized action
for $G_{\mu\nu}$ should be recovered determines the action uniquely. This form
of the action is valid in all complex dimensions $d,$ however, when $d=4,$ we
can write%
\[
X^{\overline{\kappa}\overline{\lambda}\overline{\sigma}\mu\nu\rho}=-\frac
{1}{g}\epsilon^{\overline{\kappa}\overline{\lambda}\overline{\sigma
}\,\overline{\eta}}\epsilon^{\mu\nu\rho\tau}g_{\tau\overline{\eta}},
\]
and the action takes the very simple form%
\[
I=-\frac{1}{2}%
%TCIMACRO{\dint \limits_{M}}%
%BeginExpansion
{\displaystyle\int\limits_{M}}
%EndExpansion
d^{4}zd^{4}\overline{z}\epsilon^{\overline{\kappa}\overline{\lambda}%
\overline{\sigma}\,\overline{\eta}}\epsilon^{\mu\nu\rho\tau}g_{\tau
\overline{\eta}}\partial_{\mu}g_{\nu\overline{\sigma}}\partial_{\overline
{\kappa}}g_{\rho\overline{\lambda}}.
\]
The above expression has the advantage that the action is a function of the
metric $g_{\mu\overline{\nu}}$ and there is no need to introduce the inverse
metric $g^{\overline{\nu}\mu}.$ This suggests that the action could be
expressed in terms of the K\"{a}hler form $F.$ Indeed, we can write
\[
I=\frac{i}{2}%
%TCIMACRO{\dint \limits_{M}}%
%BeginExpansion
{\displaystyle\int\limits_{M}}
%EndExpansion
F\wedge\partial F\wedge\overline{\partial}F.
\]
The equations of motion are given by
\[
\epsilon^{\overline{\kappa}\overline{\lambda}\overline{\sigma}\,\overline
{\eta}}\epsilon^{\mu\nu\rho\tau}\left(  g_{\nu\overline{\sigma}}\partial_{\mu
}\partial_{\overline{\kappa}}g_{\rho\overline{\lambda}}+\frac{1}{2}%
\partial_{\mu}g_{\nu\overline{\sigma}}\partial_{\overline{\kappa}}%
g_{\rho\overline{\lambda}}\right)  =0.
\]
Notice that the above equations are trivially satisfied when the metric
$g_{\mu\overline{\nu}}$ is K\"{a}hler%
\[
\partial_{\mu}g_{\nu\overline{\rho}}=\partial_{\nu}g_{\mu\overline{\rho}%
},\qquad\partial_{\overline{\sigma}}g_{\nu\overline{\rho}}=\partial
_{\overline{\rho}}g_{\nu\overline{\sigma}},
\]
where these conditions are locally equivalent to $g_{\mu\overline{\nu}%
}=\partial_{\mu}\partial_{\overline{\nu}}K$ for some scalar function $K.$

\section{Four dimensional limit with vanishing imaginary part}

To study the spectrum of the action we have to assume that although the
coordinates are complex, the imaginary parts are small in low-energy
experiments. The action is a function of the fields $G_{\mu\nu}\left(
x,y\right)  $ and $B_{\mu\nu}\left(  x,y\right)  $ which depend continuously
on the coordinates $y^{\mu}$ implying a continuos spectrum with an infinite
number of fields depending on $x^{\mu}.$ To obtain a discrete spectrum a
certain physical assumption should be made that forces the imaginary
coordinates to be small. One way, suggested by Witten, \cite{Witten} is to
suppress the imaginary parts by constructing an orbifold space $M^{\prime
}=M/G$ where $G$ is the group of imaginary shifts%
\[
z^{\mu}\rightarrow z^{\mu}+i(2\pi k^{\mu}),
\]
where $k^{\mu}$ are real. To maintain invariance under general coordinate
transformation we must have $k^{\mu}\left(  x,y\right)  .$ It is not easy,
however, to deal with such an orbifold in field theoretic considerations. To
determine what is needed we proceed by first expressing the full action in
terms of the fields $G_{\mu\nu}\left(  x,y\right)  $ and $B_{\mu\nu}\left(
x,y\right)  $. We  write
\[
\partial_{\mu}g_{\nu\overline{\sigma}}\partial_{\overline{\kappa}}%
g_{\rho\overline{\lambda}}=\frac{1}{4}\left(  A_{\kappa\lambda\sigma\mu\nu
\rho}+iB_{\kappa\lambda\sigma\mu\nu\rho}\right)  ,
\]
where
\begin{align*}
A_{\kappa\lambda\sigma\mu\nu\rho} &  =\left(  \partial_{\mu}^{x}G_{\nu\sigma
}+\partial_{\mu}^{y}B_{\nu\sigma}\right)  \left(  \partial_{\kappa}^{x}%
G_{\rho\lambda}-\partial_{\kappa}^{y}B_{\rho\lambda}\right)  \\
&  -\left(  \partial_{\mu}^{x}B_{\nu\sigma}-\partial_{\mu}^{y}G_{\nu\sigma
}\right)  \left(  \partial_{\kappa}^{x}B_{\rho\lambda}+\partial_{\kappa}%
^{y}G_{\rho\lambda}\right)  ,\\
B_{\kappa\lambda\sigma\mu\nu\rho} &  =\left(  \partial_{\mu}^{x}G_{\nu\sigma
}+\partial_{\mu}^{y}B_{\nu\sigma}\right)  \left(  \partial_{\kappa}^{x}%
B_{\rho\lambda}+\partial_{\kappa}^{y}G_{\rho\lambda}\right)  \\
&  +\left(  \partial_{\mu}^{x}B_{\nu\sigma}-\partial_{\mu}^{y}G_{\nu\sigma
}\right)  \left(  \partial_{\kappa}^{x}G_{\rho\lambda}-\partial_{\kappa}%
^{y}B_{\rho\lambda}\right)  .
\end{align*}
The equations of motion split into real and imaginary parts. These are given
by%
\begin{align*}
0 &  =\epsilon^{\kappa\lambda\sigma\eta}\epsilon^{\mu\nu\rho\tau}\left(
G_{\nu\sigma}\left(  \left(  \partial_{\mu}^{x}\partial_{\kappa}^{x}%
+\partial_{\mu}^{y}\partial_{\kappa}^{y}\right)  G_{\rho\lambda}-\left(
\partial_{\mu}^{x}\partial_{\kappa}^{y}-\partial_{\mu}^{y}\partial_{\kappa
}^{x}\right)  B_{\rho\lambda}\right)  \right.  \\
&  \hspace{0.6in}-B_{\nu\sigma}\left(  \left(  \partial_{\mu}^{x}%
\partial_{\kappa}^{y}-\partial_{\mu}^{y}\partial_{\kappa}^{x}\right)
G_{\rho\lambda}+\left(  \partial_{\mu}^{x}\partial_{\kappa}^{x}+\partial_{\mu
}^{y}\partial_{\kappa}^{y}\right)  B_{\rho\lambda}\right)  \\
&  \hspace{0.6in}\left.  +\frac{1}{2}A_{\kappa\lambda\sigma\mu\nu\rho}\right)
,
\end{align*}%
\begin{align*}
0 &  =\epsilon^{\kappa\lambda\sigma\eta}\epsilon^{\mu\nu\rho\tau}\left(
G_{\nu\sigma}\left(  \left(  \partial_{\mu}^{x}\partial_{\kappa}^{x}%
+\partial_{\mu}^{y}\partial_{\kappa}^{y}\right)  B_{\rho\lambda}+\left(
\partial_{\mu}^{x}\partial_{\kappa}^{y}-\partial_{\mu}^{y}\partial_{\kappa
}^{x}\right)  G_{\rho\lambda}\right)  \right.  \\
&  \hspace{0.6in}+B_{\nu\sigma}\left(  \left(  \partial_{\mu}^{x}%
\partial_{\kappa}^{x}+\partial_{\mu}^{y}\partial_{\kappa}^{y}\right)
G_{\rho\lambda}-\left(  \partial_{\mu}^{x}\partial_{\kappa}^{y}-\partial_{\mu
}^{y}\partial_{\kappa}^{x}\right)  B_{\rho\lambda}\right)  \\
&  \hspace{0.6in}\left.  +\frac{1}{2}B_{\kappa\lambda\sigma\mu\nu\rho}\right)
.
\end{align*}
We are interested in evaluating this action and equations of motion for small
values of the imaginary coordinates $y^{\mu}.$ The above expressions contain
terms which are at most quadratic in $\partial_{\mu}^{y}$ derivatives, it is
then enough to expand the fields to second order in $y^{\mu}$ and take the
limit $y\rightarrow0.$ We therefore write
\begin{align*}
G_{\mu\nu}\left(  x,y\right)   &  =G_{\mu\nu}(x)+G_{\mu\nu\rho}(x)y^{\rho
}+\frac{1}{2}G_{\mu\nu\rho\sigma}\left(  x\right)  y^{\rho}y^{\sigma}%
+O(y^{3}),\\
B_{\mu\nu}\left(  x,y\right)   &  =B_{\mu\nu}(x)+B_{\mu\nu\rho}(x)y^{\rho
}+\frac{1}{2}B_{\mu\nu\rho\sigma}\left(  x\right)  y^{\rho}y^{\sigma}%
+O(y^{3}).
\end{align*}
What is needed is a principle that determines the fields $G_{\mu\nu\rho}(x)$,
$B_{\mu\nu\rho}(x),G_{\mu\nu\rho\sigma}(x)$ and $B_{\mu\nu\rho\sigma}(x)$ and
all higher terms as functions of  $G_{\mu\nu}(x)$, $B_{\mu\nu}(x).$ For our
purposes it will be enough to determine the expansions only to second order.
This can be achieved by imposing boundary conditions in the limit
$y\rightarrow0$ on the first and second derivatives of the Hermitian metric.
The invariances of the string action given in the introduction suggests that
the equations of motion in the $y\rightarrow0$ limit reproduce the low-energy
limit of the string equations%
\begin{align*}
0 &  =G^{\eta\tau}\left(  R\left(  G\right)  +\frac{1}{6}H_{\mu\nu\rho}%
H^{\mu\nu\rho}\right)  -2\left(  R^{\eta\tau}\left(  G\right)  +\frac{1}%
{4}H_{\hspace{0.04in}\nu\rho}^{\eta}H^{\tau\nu\rho}\right)  ,\\
0 &  =\nabla^{\mu\left(  G\right)  }H_{\mu\eta\tau},
\end{align*}
In the absence of a principle that reduces the continuos spectrum, we shall
impose the boundary conditions on the Hermitian metric $g_{\mu\overline{\nu}%
}\left(  x,y\right)  $ to be such that
\begin{align*}
T_{\mu\nu\overline{\rho}}|_{y\rightarrow0}  & =2iB_{\mu\nu,\rho}\left(
x\right)  ,\\
\left[  R_{\mu\overline{\sigma}\kappa\overline{\lambda}}-R_{\kappa
\overline{\sigma}\mu\overline{\lambda}}\right]  _{y\rightarrow0}  & =-2\left(
R_{\mu\kappa\sigma\lambda}\left(  G\right)  +i\left(  \nabla_{\lambda}%
^{G}H_{\mu\kappa\sigma}-\nabla_{\sigma}^{G}H_{\mu\kappa\lambda}\right)
\right)  .
\end{align*}
The solution of the torsion constraint gives, to lowest orders,
\begin{align*}
G_{\mu\nu\rho}\left(  x\right)    & =\partial_{\nu}B_{\mu\rho}\left(
x\right)  +\partial_{\mu}B_{\nu\rho}\left(  x\right)  ,\\
B_{\mu\nu\rho}\left(  x\right)    & =-G_{\mu\rho,\nu}\left(  x\right)
+G_{\nu\rho,\mu}\left(  x\right)  ,
\end{align*}
where all derivatives are with respect to $x^{\mu}.$ Substituting these into
the curvature constraints yield%
\begin{align*}
G_{\mu\sigma\kappa\lambda}\left(  x\right)    & =\partial_{\sigma}%
\partial_{\lambda}G_{\mu\kappa}\left(  x\right)  +\partial_{\mu}%
\partial_{\lambda}G_{\sigma\kappa}\left(  x\right)  +\partial_{\sigma}%
\partial_{\kappa}G_{\mu\lambda}\left(  x\right)  \\
& +\partial_{\mu}\partial_{\kappa}G_{\sigma\lambda}\left(  x\right)
-\partial_{\kappa}\partial_{\lambda}G_{\mu\sigma}\left(  x\right)  +O\left(
\partial G,\partial B\right)  ,\\
B_{\mu\sigma\kappa\lambda}\left(  x\right)    & =\partial_{\sigma}%
\partial_{\lambda}B_{\mu\kappa}\left(  x\right)  -\partial_{\mu}%
\partial_{\lambda}B_{\sigma\kappa}\left(  x\right)  +\partial_{\sigma}%
\partial_{\kappa}B_{\mu\lambda}\left(  x\right)  \\
& -\partial_{\mu}\partial_{\kappa}B_{\sigma\lambda}\left(  x\right)
-\partial_{\kappa}\partial_{\lambda}B_{\mu\sigma}\left(  x\right)  +O\left(
\partial G,\partial B\right)  ,
\end{align*}
where $O\left(  \partial G,\partial B\right)  $ are terms of second order. To
write the equations of motion in component form, we substitute the $G_{\mu\nu
}(x,y)$ and $B_{\mu\nu}(x,y)$ expansions into $A_{\kappa\lambda\sigma\mu
\nu\rho}$ and $B_{\kappa\lambda\sigma\mu\nu\rho}$ using the above solutions to
obtain
\begin{align*}
A_{\kappa\lambda\sigma\mu\nu\rho} &  =\Gamma_{\mu\nu\sigma}\left(  G\right)
\Gamma_{\kappa\lambda\rho}\left(  G\right)  \\
&  -\left(  \partial_{\mu}B_{\nu\sigma}+\partial_{\sigma}B_{\mu\nu}%
-\partial_{\nu}B_{\sigma\mu}\right)  \left(  \partial_{\kappa}B_{\rho\lambda
}+\partial_{\lambda}B_{\rho\kappa}+\partial_{\rho}B_{\rho\kappa}\right)
+O(y),\\
B_{\kappa\lambda\sigma\mu\nu\rho} &  =\Gamma_{\mu\nu\sigma}\left(  G\right)
\left(  \partial_{\kappa}B_{\rho\lambda}+\partial_{\lambda}B_{\rho\kappa
}+\partial_{\rho}B_{\rho\kappa}\right)  \\
&  +\left(  \partial_{\mu}B_{\nu\sigma}+\partial_{\sigma}B_{\mu\nu}%
-\partial_{\nu}B_{\sigma\mu}\right)  \Gamma_{\kappa\lambda\rho}\left(
G\right)  +O(y),
\end{align*}
where
\[
\Gamma_{\mu\nu\sigma}\left(  G\right)  =\partial_{\nu}G_{\mu\sigma}%
+\partial_{\mu}G_{\nu\sigma}-\partial_{\sigma}G_{\mu\nu}.
\]
In terms of components, the equations of motion take the form%
\begin{align*}
0 &  =\epsilon^{\kappa\lambda\sigma\eta}\epsilon^{\mu\nu\rho\tau}\left(
G_{\nu\sigma}\left(  \partial_{\mu}\partial_{\kappa}G_{\rho\lambda}%
+G_{\rho\lambda\mu\kappa}-\partial_{\mu}B_{\rho\lambda\kappa}+\partial
_{\kappa}B_{\rho\lambda\mu}\right)  \right.  \\
&  \hspace{0.6in}\qquad-B_{\nu\sigma}\left(  \partial_{\mu}\partial_{\kappa
}B_{\rho\lambda}+B_{\rho\lambda\mu\kappa}+\partial_{\mu}G_{\rho\lambda\kappa
}-\partial_{\kappa}G_{\rho\lambda\mu}\right)  \\
&  \hspace{1in}\left.  +\frac{1}{2}A_{\kappa\lambda\sigma\mu\nu\rho}\right)
,\\
0 &  =\epsilon^{\kappa\lambda\sigma\eta}\epsilon^{\mu\nu\rho\tau}\left(
G_{\nu\sigma}\left(  \partial_{\mu}\partial_{\kappa}B_{\rho\lambda}%
+B_{\rho\lambda\mu\kappa}+\partial_{\mu}G_{\rho\lambda\kappa}-\partial
_{\kappa}G_{\rho\lambda\mu}\right)  \right.  \\
&  \hspace{1in}-B_{\nu\sigma}\left(  \left(  \partial_{\mu}\partial_{\kappa
}G_{\rho\lambda}+G_{\rho\lambda\mu\kappa}-\partial_{\mu}B_{\rho\lambda\kappa
}+\partial_{\kappa}B_{\rho\lambda\mu}\right)  \right)  \\
&  \hspace{1in}\left.  +\frac{1}{2}B_{\kappa\lambda\sigma\mu\nu\rho}\right)  .
\end{align*}
After substituting the solutions of the constraints these take the form
\begin{align*}
\epsilon^{\kappa\lambda\sigma\eta}\epsilon^{\mu\nu\rho\tau}\left(
G_{\nu\sigma}R_{\mu\lambda\rho\kappa}\left(  G\right)  -\frac{1}{4}%
\partial_{\sigma}B_{\mu\nu}\partial_{\rho}B_{\kappa\lambda}-2B_{\nu\sigma
}\partial_{\lambda}\partial_{\mu}B_{\rho\kappa}\right)    & =0,\\
\epsilon^{\kappa\lambda\sigma\eta}\epsilon^{\mu\nu\rho\tau}\left(
G_{\nu\sigma}\left(  \partial_{\lambda}\partial_{\mu}B_{\rho\kappa}\right)
-B_{\nu\sigma}R_{\mu\lambda\rho\kappa}\left(  G\right)  \right)    & =0.
\end{align*}
Using the identity
\[
\epsilon^{\kappa\lambda\sigma\eta}\epsilon^{\mu\nu\rho\tau}G_{\nu\sigma}%
=6\det\left(  G_{\mu\nu}\right)  G^{\mu\left[  \kappa\right.  }G^{\lambda
|\rho}G^{\left.  \eta\right]  \tau},
\]
these equations reduce to the correct equations of motion, up to terms of the
form $O\left(  \partial G,\partial B\right)  $ which were neglected in the derivation.

\section{Conclusions}

In this work we have investigated the structure of a complexified space-time.
The geometry is taken to be that of a Hermitian manifold with complex metric
given by $g_{\mu\overline{\nu}}\left(  z,\overline{z}\right)  =G_{\mu\nu
}\left(  x,y\right)  +iB_{\mu\nu}(x,y).$ After studying the properties of
Hermitian geometry, we find that there is a unique action, up to boundary
terms, that gives the correct linearized kinetic energies for $G_{\mu\nu}(x)$
and $B_{\mu\nu}\left(  x\right)  $ in the limit when the metric is restricted
to depend only the variables $x^{\mu}.$ The unique action is of the
Chern-Simons type when expressed in terms of the K\"{a}hler form. We have
shown that the diffeomorphism invariance in the complex coordinates protect
both fields $G_{\mu\nu}(x)$ and $B_{\mu\nu}\left(  x\right)  $ keeping them
massless.  The physical requirement that the imaginary parts of the
coordinates are small at low energies, must be imposed in such a way as to
reduce the continuos spectrum of $G_{\mu\nu}\left(  x,y\right)  $ and
$B_{\mu\nu}(x,y)$ to a discrete spectrum. In the absence of information about
the spectrum arising at high energies where the imaginary coordinates are
expected to play a role, it is enough for our purposes to impose conditions on
first and second derivatives of the Hermitian metric, which allows us to solve
for the lowest order terms in the expansion in terms of $y^{\mu}.$ These
constraints are imposed on the torsion and curvature of the Hermitian geometry
in the limit $y^{\mu}\rightarrow0.$ We have solved the constraints and shown
that the equations of motion for the Hermitian metric results in the
low-energy string equations in the limit $y^{\mu}\rightarrow0.$ The results
obtained so far, give circumstantial evidence that space-time might be
enlarged to become complex.  Much more work is needed to determine the
principle that restricts the form of the hermitian metric to give a discrete
spectrum and fixes the dependence on the imaginary coordinates to all orders.
This will be necessary in order to understand the contributions of the
imaginary parts of the coordinates at high energies. One would expect that
$B_{\mu\nu}\left(  x\right)  $ would also enter in the higher order terms of
the action and not only through their derivatives, in analogy with the field
$G_{\mu\nu}\left(  x\right)  .$

\section{Acknowledgment}

I\ would like to thank Jean-Pierre Bourguignon for pointing reference
\cite{Gaud} to me. Research supported in part by the National Science
Foundation under Grant No. Phys-0313416.

\end{document}